# Optical Widefield Nuclear Magnetic Resonance Microscopy

Karl D. Briegel[1,2], Nick R. von Grafenstein[1,2]†, Julia C. Draeger[1,2]†, Peter Blümler[3], Robin D. Allert[1,2], Dominik B. Bucher[1,2]*

[1] Technical University of Munich, TUM School of Natural Sciences, Chemistry Department, Technical University of Munich, Lichtenbergstraße 4,85748 Garching bei München, Germany
[2] Munich Center for Quantum Science and Technology (MCQST), Schellingstr. 4, D-80799
[3] University of Mainz, Institute of Physics, Staudingerweg 7, 5528 Mainz, Germany

†These authors contributed equally to this work
*Corresponding author: dominik.bucher@tum.de

**Abstract**

**Microscopy enables detailed visualization and understanding of minute structures or processes. While cameras have significantly advanced optical, infrared, and electron microscopy, imaging nuclear magnetic resonance (NMR) signals on a camera has remained elusive. Here, we employ nitrogen-vacancy (NV) centers in diamond as a quantum sensor, which converts NMR signals into optical signals that are subsequently captured by a high-speed camera. Unlike traditional magnetic resonance imaging (MRI), our method records the NMR signal over a wide field of view in real space. We demonstrate that our optical widefield NMR microscopy (OMRM) can image NMR signals in microfluidic structures with a ~10 μm resolution across a ~235 × 150 μm² area. Crucially, each camera pixel records an NMR spectrum providing multicomponent information about the signal's amplitude, phase, local magnetic field strengths, and gradients. The fusion of optical microscopy and NMR techniques enables multifaceted imaging applications in the physical and life sciences.**

## Introduction

Cameras have revolutionized microscopy, spanning visible light, fluorescence, infrared, and electron microscopy. In contrast to scanning methods, widefield detection with cameras increases speed and throughput by performing parallel measurements. However, the use of cameras in nuclear magnetic resonance/imaging (NMR/MRI) techniques has been hampered by their inability to detect magnetic resonance signals directly. In this study, we use nitrogen-vacancy (NV) centers in diamond as microsensors to convert local NMR signals into optical signals (*1–8*) that can be detected by a camera in real space, allowing parallel widefield imaging. This approach differs significantly from conventional MRI (*9*), which relies on encoding spatial information via magnetic field gradients. We use a high-speed camera to stroboscopically interrogate the NMR signal encoded in the fluorescence intensity of an NV-doped diamond chip. We first establish a high-speed camera readout for a quantum sensing protocol designed for NMR signal detection. Subsequently, we apply this technique to image NMR signals from within the microstructures of a microfluidic chip, achieving a spatial resolution of ~10 μm over a field of view of ~235 × 150 μm². Each pixel within the captured images provides information on the local amplitude and phase of the NMR signal, exhibiting a strong dependence on the local geometry, as validated by our simulations. In addition, the NMR signal provides information about local magnetic fields and gradients. Thus, our optical widefield nuclear magnetic resonance microscopy (OMRM) effectively bridges the gap between optical microscopy and information-rich magnetic resonance methods.

## Results

**Principle of operation and experimental design.** Nuclear spins (e.g., $^1$H nuclei) can be excited by a radiofrequency (RF) pulse, causing them to precess around an applied magnetic field $B_0$ and emit an RF/NMR signal at their Larmor frequency (Fig. 1A). Instead of detecting the NMR signal inductively, we place the sample on an NV center doped diamond chip. By coherent microwave (MW) control of the NV centers, we couple their electronic spin states to the NMR signal (*10, 11*). The radius of the dominant detection volume which determines the spatial resolution, roughly corresponds to twice the average depth of the NV center ensemble from the diamond surface (Fig. 1B, (*3, 12*)). Since the NV center's fluorescence intensity is spin-state dependent (*13, 14*), it converts the NMR signal from a magnetic to an optical signal (Section S1). By imaging the spatial fluorescence intensity of the NV diamond on a camera, we can record the transcoded NMR signal and thus obtain a spatially distributed map of local NMR spectra (Fig. 1C). Thus, the spectra are recorded in parallel for each camera pixel and do not require further encoding steps or magnetic field gradients to obtain spatial information.

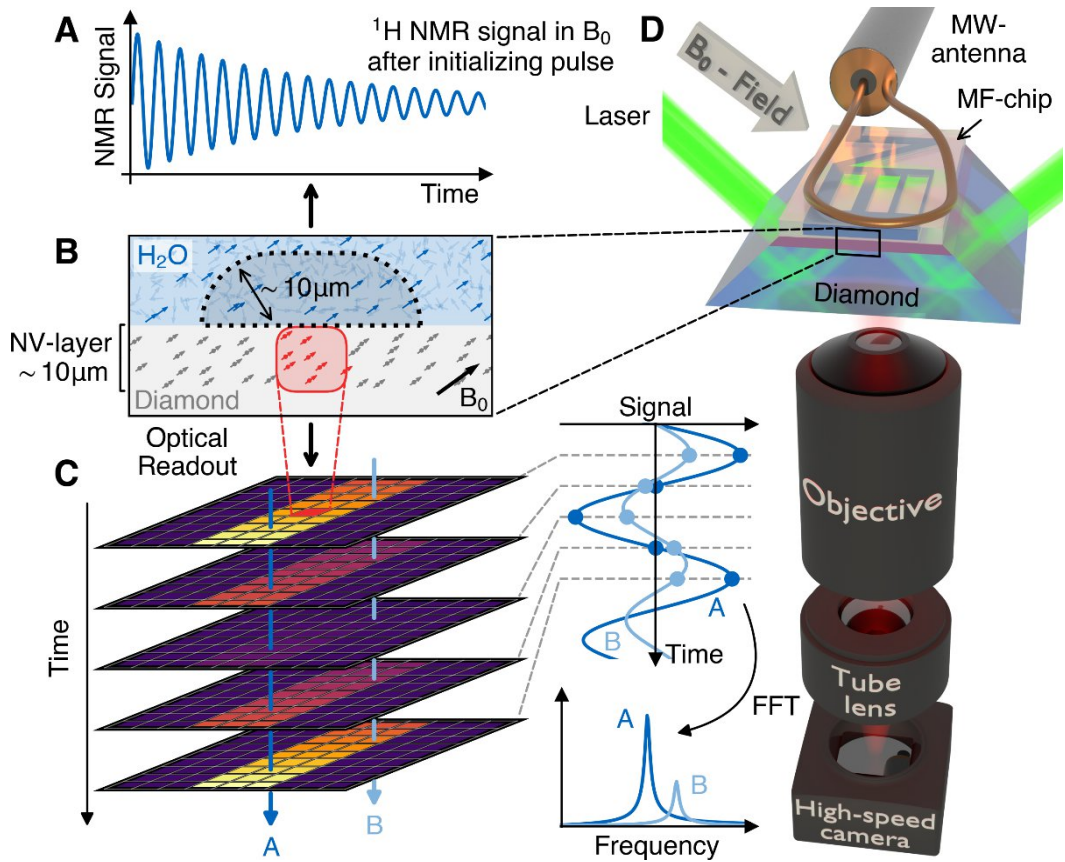

**Fig. 1. Basic principle of optical widefield magnetic resonance microscopy.** (**A**) NMR signals are caused by the precession of nuclear spins (e.g., $^1$H) around an applied magnetic field $B_0$ after excitation by a radiofrequency pulse. (**B**) The NMR signal is detected optically using NV centers as local magnetometers in the diamond. We use a ~10 μm thick NV-doped diamond layer, which sets the spatial resolution to a similar length scale. (**C**) The fluorescence intensity encoded NMR signal is imaged with a high speed streaming camera. Each pixel time trace, corresponding to a location on the diamond, can be Fourier transformed (FFT) into a frequency spectrum. (**D**) Experimental setup: The NV layer of the diamond is optically excited at 532 nm using a total internal reflection geometry. The NMR sample is in a microfluidic (MF) structure on top of the diamond. For sensing, the NV spins are coupled to the sample spins by microwave pulses from an antenna. The resulting NV's spin state-dependent fluorescence intensity is captured by an objective and imaged with a tube lens onto a high-speed streaming camera. The probe head is positioned within a highly homogeneous magnetic field $B_0$ at ~84 mT for NMR detection. Further details can be found in M&M.

The core of the OMRM microscope is an electronic grade diamond with a ~10 μm overgrown NV-doped diamond layer, which defines the spatial resolution (Fig. 1D, Section 2). The diamond is polished into a trapezoidal shape in order to couple in the green excitation light (532 nm) in a total internal reflection geometry (*3*). We use a laser spot to illuminate the NV layer over an area of ~235 × 150 μm$^2$. The diamond is bonded into a custom-designed microfluidic glass chip (*15*). To control the NV spin state, MW pulses are delivered via a short-circuited coaxial cable placed on top of the microfluidic chip (*16*). A 3D-printed sample holder houses the microfluidic chip and mounts a pair of copper wire coils for the RF excitation of the proton sample spins. The entire probe head, made of non-magnetic materials, is mounted in a custom-built highly homogeneous permanent magnet (*17*). A high numerical aperture (0.8NA) objective with low magnetic properties and tube lens are used to collect and image the fluorescence of the NV layer onto a high-speed streaming camera. Residual excitation light is filtered with a long pass filter to retain the signal-carrying fluorescence (details in M&M). With this setup, we achieve a sensitivity of ~10-30 nT Hz$^{-1/2}$ μm$^{3/2}$ for RF signals (Section S3).

**NV quantum sensing over a wide field of view.** In all measurements, the NV-center's electron spin state is optically initialized by laser excitation. MW pulse sequences are subsequently used to manipulate the NV spin for sensing over the field of view, and the change in the spin-state of the NV is optically imaged (Fig 2A, (*11*)). In the first experiment, we determine the NV Rabi frequency, which forms the basis of the pulse sequence for detecting NMR signals. By applying progressively longer MW pulses at the NV resonance frequency, we rotate the NV spin state and extract the per-pixel spin control parameters that lead to half and full spin flips, i.e., π/2- and π-pulses (Fig. 2B). Color coding of the π-pulse duration provides a map of the MW driving field strength (Fig. 2C), which varies slightly across the field of view, due to the geometry of the MW antenna.

To detect high frequency-resolved NMR signals with NV centers, we use the coherently averaged synchronized readout (CASR) pulse sequence (Section S1, (*3*)). CASR based on a precisely timed train of π-pulses that couples the NV-center's spin to the NMR signal by matching the repetition rate of the π-pulses to half the period of the expected NMR signal frequency (dynamical decoupling sequence (*11*), Section S1) (Fig. 2D). For a given NMR signal and MW pulse train, the magnitude of the coupling is a function of the phase of the NMR signal relative to the π-pulse train. A precisely timed repetition of the coupling and readout process results in a time-varying coupling amplitude at an aliased frequency of the original NMR signal, which can be detected optically using the NV photoluminescence (Fig. 1B and Fig. 2E). In previously published NV-based magnetic imaging studies, the experimental procedure allows repetition and averaging over individual sweep parameters, such as the MW frequency in quantum diamond microscopy or the pulse duration in Rabi experiments (*18–25*). However, when the CASR pulse sequence is used for NMR signal detection, real-time readout (~5,000-20,000 frames/s) becomes essential (Section S4, (*26*, *27*)). This is due to the inherent synchronization requirements between the CASR pulse sequence and the (decaying) NMR signal (Fig. 1C & 2D). This is trivial for a photodiode but poses technical challenges for imaging systems. Our solution is a high-speed streaming camera with a framerate of up to multiple tens of kHz (M&M, Section S4).

Secondly, we demonstrate the detection of continuous RF calibration signals from a nearby antenna emitting at ~3.56 MHz. Each pixel records a photoluminescence (PL) oscillation at the aliased RF frequency in the time domain (Fig. 2E, inset) which is Fourier transformed. Color coding of the corresponding signal-to-noise ratio (SNR) yields an image (Fig. 2F). We observe a similar spatial variation of the SNR as in the Rabi image. The two are related because well-defined π pulse durations are required for the CASR pulse sequence, and variations will reduce its sensitivity. For this reason, the field of view is currently limited by the homogeneity of the Rabi frequency distribution (MW field homogeneity, Fig. 2C & 2F), which can be mitigated in the future by using MW resonators (*10*).

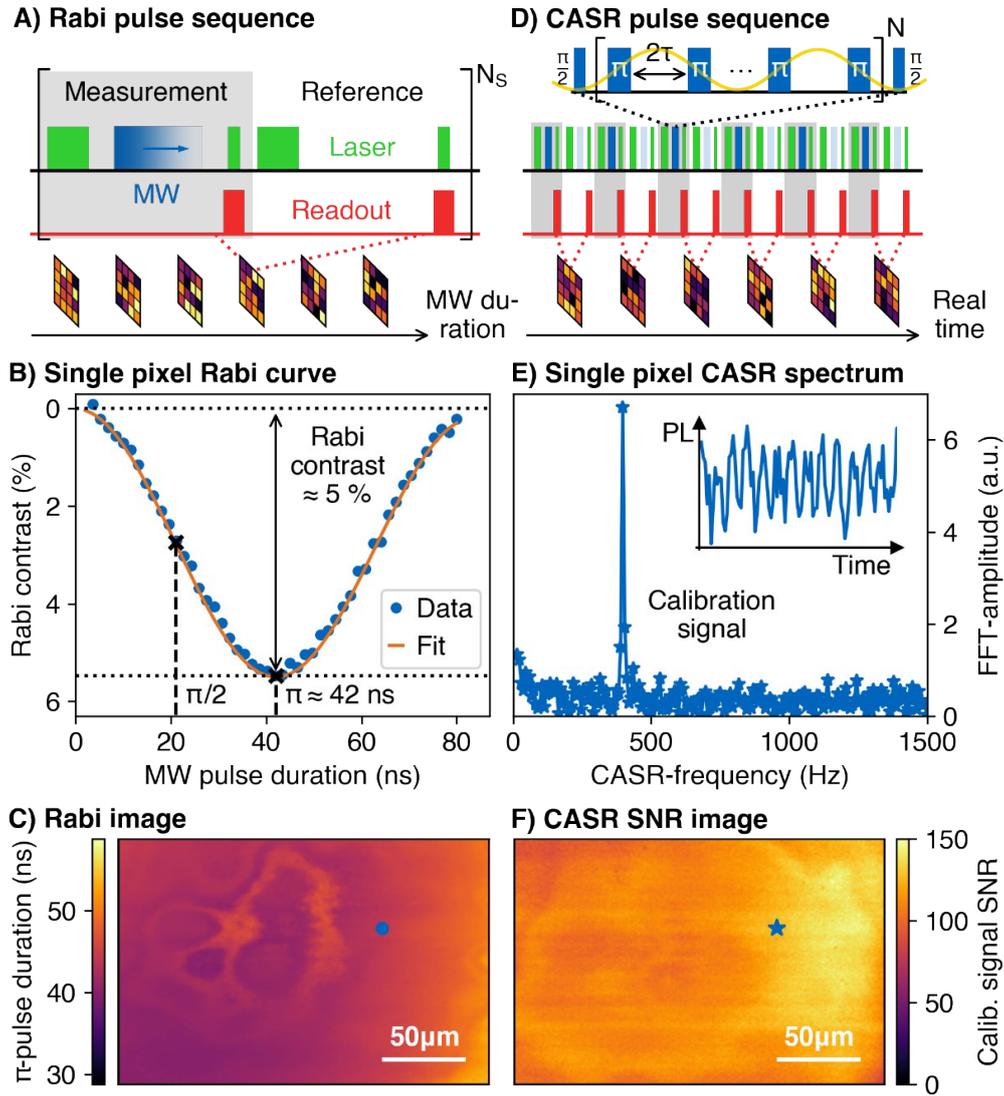

**Fig. 2. Widefield quantum sensing with a high-speed streaming camera.** The employed NV quantum sensing protocols require an initialization laser pulse (green), a microwave (MW) pulse (blue) to control the NV spin-state and a second laser pulse to read out the spin-dependent photoluminescence (PL) of the NV (red), which is captured on a camera. (**A**) For a Rabi experiment, the NV center's spin-state is observed as a function of the MW pulse duration. For each MW pulse duration, the spin state-dependent PL, as well as one reference image for common mode noise rejection, are recorded and averaged $N_s$ times. (**B**) Increasing the MW pulse duration leads to spatially resolved Rabi oscillations for each pixel. The π/2- and π-pulse durations, determined for each pixel, serve as important control parameters for the coherently averaged synchronized readout (CASR) sequence to detect NMR signals. (**C**) Image of the π pulse duration determined from the single pixel Rabi fits. The pixel corresponding to the dataset in (B) is marked with a blue dot. (**D**) Radiofrequency (RF) signals are detected stroboscopically using the CASR pulse sequence. CASR is based on blocks of MW pulses (blue) starting and ending with a π/2- and a train of N π-pulses coupling the NV centers to the detected RF field (yellow). After each sensing block, the NV-fluorescence is detected, requiring a time-resolved readout on a high-speed streaming camera. (**E**) Detection of an RF calibration signal. The recorded pixel-wise time domain data (inset) is Fourier transformed and shows a signal at the aliased CASR frequency. (**F**) Image of the signal-to-noise ratio (SNR) of the single pixel analysis of the CASR data. The pixel corresponding to the dataset in (E) is marked with a blue star.

**Optical widefield magnetic resonance microscopy (OMRM).** In the next step, we utilize the CASR protocol to detect NMR signals from a water sample contained within the microfluidic structure. We apply Overhauser hyperpolarization to increase the NMR signal by two orders of magnitude at our low magnetic fields (~84 mT, $^1$H NMR at ~3.56 MHz) (*28, 29*). An RF pulse induces the NMR signal which is detected with the CASR pulse sequence (Fig. 3A). In contrast to the RF calibration signal, the NMR signal has a limited lifetime ($T_2/T_2^*$) after the excitation. To maximize the lifetime of the NMR signal, we use a highly homogeneous magnetic field $B_0$ from a custom-designed permanent magnet and purposefully chosen non-magnetic components for the probe head (details in M&M).

In our first OMRM example, we image the NMR signals of water within the marked area of the microfluidic chip (Fig. 3B). Applying the Fourier transformation (FFT) along each pixel time domain followed by 2D median filter over the resulting frequency images yields single pixel NMR spectra (Section S5). Color coding the SNR of the NMR signal results in an OMRM image that clearly depicts the microfluidic channel filled with water (Fig. 3C, M&M). Each pixel on the camera - encoding the spatial information - contains a full NMR spectrum (Fig. 3D-E, Section S2).

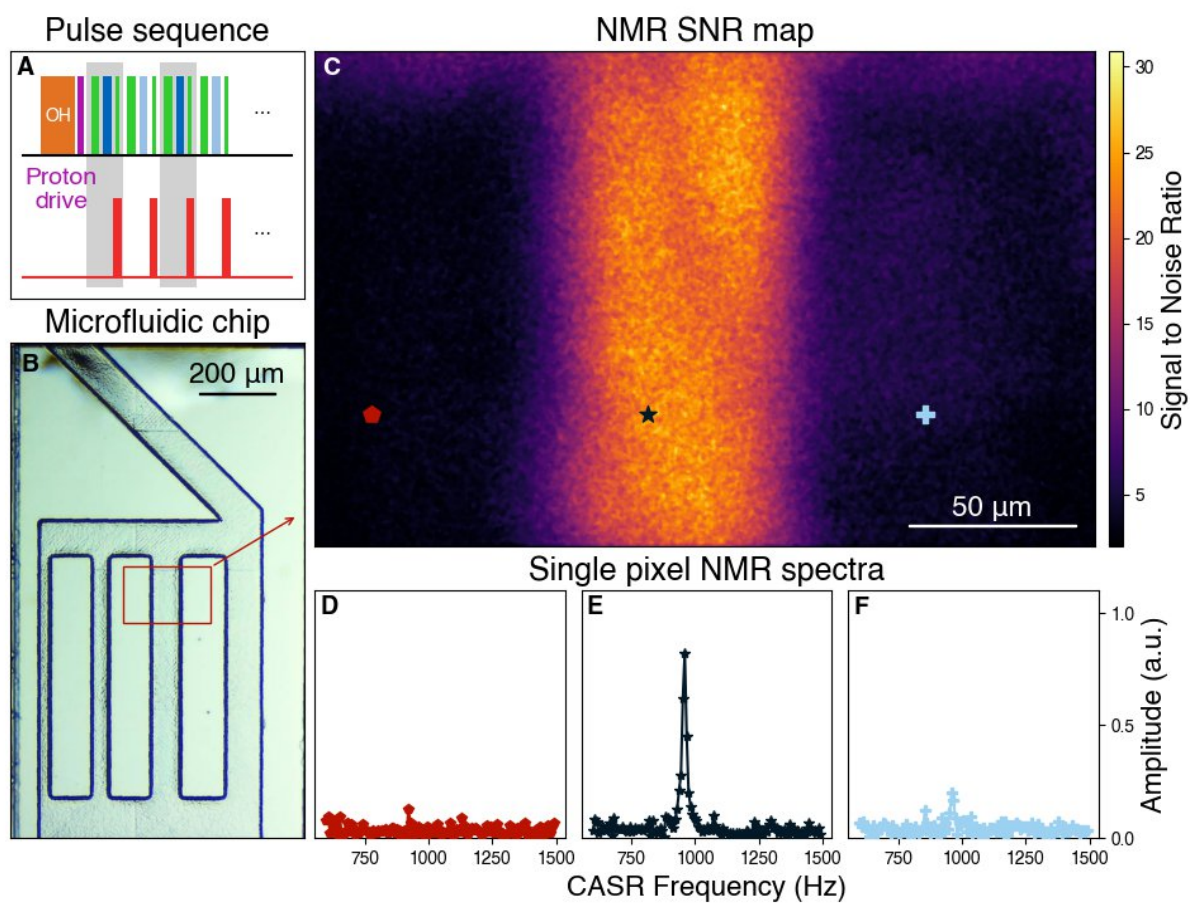

**Fig. 3. Optical widefield magnetic resonance microscopy (OMRM).** (**A**) NV-NMR pulse sequence. The (orange) Overhauser (OH) hyperpolarization and (magenta) $^1$H excitation pulses are followed by the CASR pulse sequence, which converts the NMR signal into an optical signal. (**B**) Optical microscopy image of the microfluidic chip. Our water sample is pumped and distributed through microfluidic channels through an inlet on the top part of the chip. (**C**) SNR map of the NMR spectrum obtained by Fourier transforming the single pixel NMR signals. The SNR encodes the presence of the $^1$H sample in the microfluidic chip. The data was recorded from the red region in (B). (**D-F**) Single-pixel NMR spectra at the marked locations in (C) vs CASR-frequency ($^1$H NMR frequency at ~3.56 MHz aliased to ~960 Hz). Detailed description of the data analysis can be found in Section S5.

**Multicomponent analysis of the OMRM image.** For more detailed analysis, we fit a Lorentzian function (Eq. 1) to each pixel NMR spectrum (full data analysis and complementary analysis can be found in Section S5). This yields information about the signal amplitude $S_0$, phase $\phi$, frequency $f_0$, and linewidth $\Gamma$, of the NMR signal for each pixel.

$$S(f) = S_0 \cdot e^{i\phi} \cdot \frac{\Gamma - i \cdot (f_0 - f)}{\Gamma^2 + (f_0 - f)^2} \quad \text{Eq. 1}$$

To analyze parameters other than signal size more accurately, we remove unreliable fits if their raw data SNR is below a certain threshold (Section S5). By observing the amplitude of the $^1$H NMR signal ($S_0$), we can reconstruct the channel path of the microfluidic chip (Fig. 4A & B). Due to the dipolar nature of the spin interaction and its associated symmetries, the amplitude of the NMR signal is dependent on the geometry of the NMR sample relative to the NV center (*12*). In Fig. 4A, the axis of the NV center and the magnetic field $B_0$ are orthogonal to the vertical microfluidic channel (left). This leads to an asymmetric signal amplitude across the channel width, which is confirmed by our simulations (Fig. 4B, bottom, Section S6). Shifting the field of view to a different region of the microfluidic chip results in a channel that runs at an angle of 45° relative to the NV center's axis (Fig. 4A, middle). This channel orientation changes the interaction symmetry of the NV centers, eliminating the signal amplitude asymmetry in agreement with our simulations. A final position on the microfluidic chip with a T-junction of the channels further supports this observation (Fig. 4A, right). The channel running orthogonal to the NV center's principal axis reproduces the observed asymmetry in $S_0$.

An additional parameter is the phase $\phi$ of the NMR signal, which also depends on the geometry and varies accordingly across the microfluidic channels (Fig. 4C). While the origin of this observation is the same as before, the observed gradients in $\phi$ follow the opposite trend compared to $S_0$. Fig. 4C shows the same channel sections as in Fig. 4B; however, the signal phase is symmetric around the vertical channel (Fig. 4C, left). Shifting the field of view to the diagonal section of the channel changes the interaction symmetry and introduces a phase gradient across the channel (Fig. 4C, middle). The T-junction reproduces the symmetric phase distribution around the orthogonal portion of the channel, while the parallel part exhibits a phase gradient across its width (Fig. 4C, right).

The signal frequency ($f_0$) provides information about the Larmor frequency of the sample, which can give important chemical information. Since our sample is water, which exhibits a single $^1$H resonance, $f_0$ serves as a highly accurate measure of the local magnetic field strength across the field of view (Fig. 4D). In our measurements, the observed gradient was ~6 nT/µm for all chip positions, indicating a macroscopic magnetic field gradient.

Finally, we can resolve the local NMR linewidth ($\Gamma$), which provides information about the proton relaxation time. We observe systematic changes in $\Gamma$ across the field of view, which are likely caused by local magnetic field gradients induced by susceptibility mismatches.

We note that the data had to be corrected for slow magnetic field ($B_0$) drifts caused by the temperature-dependent magnetization of the permanent magnet. To correct for this frequency drift, we store the acquired data every ~5 minutes and frequency correct the time domain NMR signal (M&M, Section S5). Therefore, the amplitude and linewidth values obtained should not be compared between different measurements (Section S5).

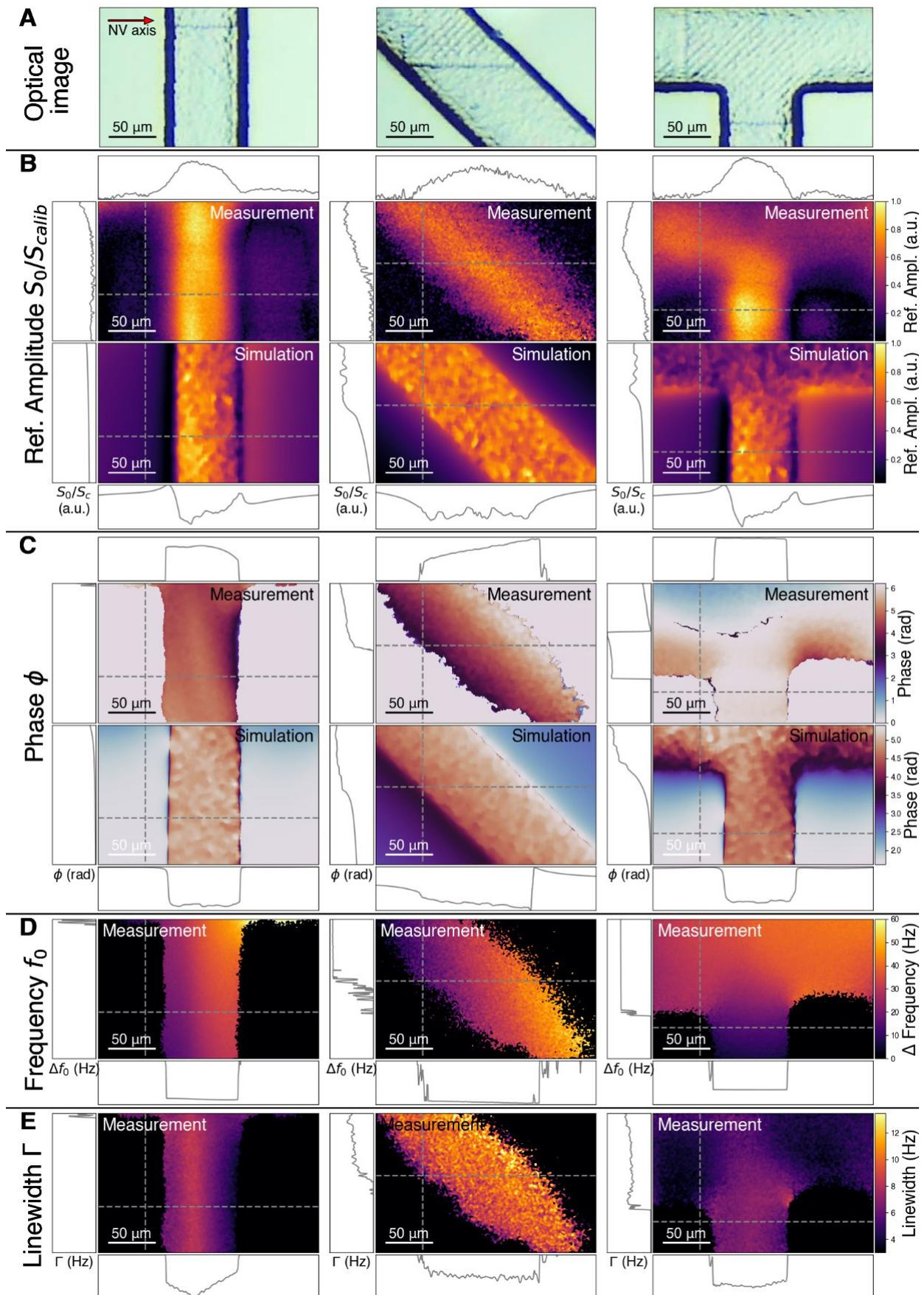

**Fig. 4. Multicomponent optical widefield NMR microscopy.** (**A**) Optical photographs depicting microfluidic chip sections that were subsequently imaged with OMRM in (B-E) (left, middle, right). (**B**)

Top row: Referenced NMR amplitude ($S_0/S_{Calib}$, Section S5) image obtained from a Lorentzian fit at three different locations on the microfluidic chip (A). Dotted lines indicate a cut through the image, plotted adjacent to them. Bottom row: Corresponding simulated amplitude maps for the same channel regions indicate an influence of the geometry on the NMR amplitude. (**C**) Top row: Fitted relative phase ($\phi$) image of the NMR signal. Bottom row: Corresponding simulated phases of the NMR signals indicate an influence of the geometry on the NMR phase. (**D**) Image of the NMR signal frequencies ($f_0$) across the field of view indicate a macroscopic gradient in $B_0$. (**E**) Linewidth ($\Gamma$) of the NMR signal indicates local magnetic field gradients. The full data processing can be found in Section S5.

**Conclusion and Outlook**

These measurements constitute a demonstration of microscale optical widefield NMR microscopy with rich information content. Presently, our spatial resolution is limited by the NV layer thickness of ~10 µm (Section S2). The spatial resolution of the OMRM technique is technically limited by the optical diffraction limit and the NMR sensitivity, where the latter scales proportionally with the number of NV centers per pixel (Section S7). For the current setup and spatial resolution, we obtain a sensitivity of ~10-30 nT Hz$^{-1/2}$ µm$^{3/2}$ with vast optimization possibilities in the future (detailed discussion in Section S8). With these improvements, achieving single micrometer spatial resolution will be feasible.

In summary, our results demonstrate the fusion of optical widefield microscopy with NMR spectroscopy, synergistically combining the strengths of each technique: the high-speed and high spatial resolution of optical imaging on a camera, together with the unique and comprehensive chemical and physical insights provided by magnetic resonance methods. In the future, we expect to increase the wealth of information by incorporating ($T_1$ and $T_2$) relaxation, diffusion (*30*) and (multinuclear) high spectral resolution spectroscopy for structural molecular information (*3*, *6*). As a result, our methodology will provide unprecedented insights into the microscopic world, with diverse applications ranging from (metabolomic) analysis of single cells (*31*) or tissues to high-throughput NMR spectroscopy and materials science (*32*).

**Acknowledgments**

**Funding.** This project has been funded by the Bayerisches Staatministerium für Wissenschaft und Kunst through project IQSense via the Munich Quantum Valley (MQV), by the Federal Ministry of Education and Research (BMBF) as part of the VIP+ validation funding program (03VP10350) and the European Research Council (ERC) under the European Union's Horizon 2020 research and innovation programme (Grant Agreement No. 948049). The authors acknowledge further support by the DFG under Germany's Excellence Strategy–EXC 2089/1-390776260 and the EXC-2111 390814868.

**Author Contributions.** D.B.B. conceived and supervised the study. K.D.B. wrote the code for hardware control, pulse sequence generation, measurement execution, and data processing. K.D.B., J.C.D. and N.R.v.G. designed and built the experiment. K.D.B., J.C.D. conducted the experiments. R.D.A. contributed to various important parts of the experimental setup. P.B. and N.R.v.G. designed, built and characterized the permanent magnet. K.D.B. performed the simulations. K.D.B., J.C.D. and D.B.B. analyzed and discussed the data and wrote the manuscript with inputs from all authors.

**List of Supplementary Materials.**

Materials and Methods

Supplementary Text S1-S8

Figures S1 to S12

Tables S1 to S3